# DEVELOPMENT AND PERFORMANCE OF RFD CRAB CAVITY PROTOTYPES FOR HL-LHC AUP*

L. Ristori†, P. Berrutti, M. Narduzzi, Fermilab, USA
J. Delayen, S. de Silva, Old Dominiun University, USA
Z. Li, A. Ratti SLAC, USA; N. Huque, A. Castilla Jlab, USA

## Abstract

The US will be contributing to the HL-LHC upgrade at CERN with the fabrication and qualification of RFD crabbing cavities in the framework of the HL-LHC Accelerator Upgrade Project (AUP) managed by Fermilab. AUP received Critical Decision 3 (CD-3) approval by DOE in December 2020 launching the project into the production phase. The electro-magnetic design of the cavity was inherited from the LHC Accelerator Research Program (LARP) but needed to be revised to meet new project requirements and to prevent issues encountered during beam tests performed at CERN in the R&D phase. Two prototype cavities were manufactured in industry and cold tested. Challenges specific to the RFD cavity were the stringent interface tolerances, the pole symmetry, and the higher-order-mode impedance spectrum. Chemical processing and heat treatments were performed initially at FNAL/ANL and are now being transferred to industry for the production phase. HOM dampers are manufactured and validated by JLAB. A summary of cold test results with and without HOM dampers is presented.

## RF DESIGN

The RFD cavity utilizes the TE-11-like dipole mode to achieve crabbing of particle bunches. The cavity is consisted of two ridged deflecting poles in a square-shaped tank. The transverse shape was optimized to fit into the limited space available at the LHC beam line as shown in Figure 1.

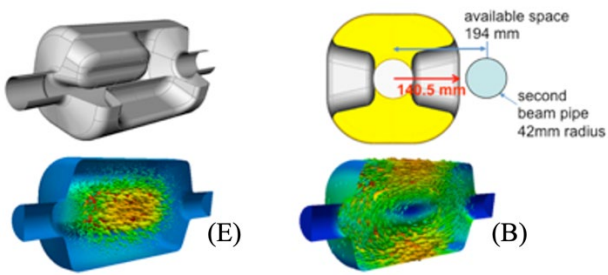

Figure 1. RFD cavity geometry (above left and right) and the electric and magnetic field patterns of its TE11-like deflecting mode (below left and below right).

With the transverse dimension being constrained, the cavity length and the rounding radius of the outer perimeter are the main parameters used to tune the cavity to the 400 MHz of design frequency. The pole length and the end gap are optimized to achieve a maximal shunt impedance. Elliptical rounding around the side corners of the poles and at around the pole tips were employed to minimize the peak surface fields. A curved profile is introduced to the gap surface to compensate the higher multipole fields due to the 3D geometry, especially the dominant sextupole term $b_3$. Design parameters of the RFD cavity are shown in Table 1.

Table 1. RFD cavity parameters, including Higher-Order Modes (HOM).

| Operating Mode | TE11 |
| --- | --- |
| Frequency (MHz) | 400.79 |
| Lowest dipole HOM (MHz) | 636 |
| Lowest acc HOM | 699 |
| High R/Q acc HOM | 752.2 |
| Iris aperture (diameter) (mm) | 84 |
| Transverse dimension (mm) | 281 |
| Vertical dimension (mm) | 281 |
| $R_T$ (ohm/cavity) | 431 |
| $V_T$ (MV/cavity) | 3.34 |
| Bs (mT) | 55.1 |
| Es (MV/m) | 35.0 |
| Multipole component b3 (mT/m$^2$/10MV) | 429 |

### Higher-Order Mode Couplers

In the RFD design, there is no lower order mode. The deflecting mode is the lowest mode of the cavity. The lowest HOM frequency is more than 230 MHz above the operating mode. This makes it straight forward to utilize a high-pass filter approach in the HOM coupler design for wakefield damping. Figure 2 shows the RFD cavity with the HOM and FPC couplers.

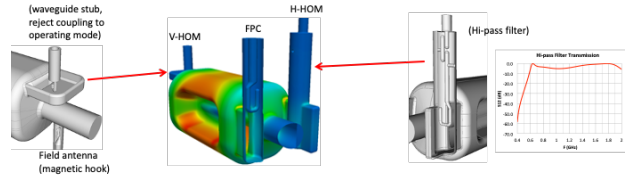

Figure 2. RFD cavity with HOM and FPC couplers. The H-HOM high-pass filter transmission curve is shown on the right.

The couplings are all through the end plate where both the electric and magnetic fields of the operating mode are low. As such, the coupler opening does not introduce local high surface fields, and the RF heating in the couplers is also of low level. The horizontal HOM (H-HOM) coupler requires a high-pass filter to reject the operating mode. The vertical HOM (V-HOM) coupler does not couple to the operating mode due to the anti-symmetry of the field thus no filter is needed. The HOM impedance of the RFD design is shown in Figure 3. The blue and red dashed lines are the upper limits of the design requirements for the transverse and longitudinal impedances respectively. Both the Fundamental Power Coupler (FPC) and the field pickup couplers utilize hooks to enhance coupling, which minimizes the side effects of RF heating on the couplers and beam direct coupling.

___
* Work supported by Fermi Research Alliance, LLC under Contract No. DEAC02-07CH11359 with the United States Department of Energy
† leoristo@fnal.gov

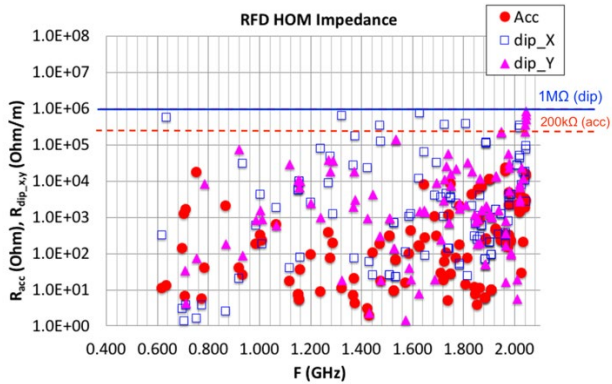

Figure 3.   The HOM spectrum of the RFD cavity. The blue and red dashed lines are the upper limits of the design requirements for the transverse and longitudinal impedances respectively.

## CAVITY FABRICATION

Cavities are under procurement at Zanon Research & Innovation (ZRI). Two cavity prototypes have been completed (see Figure 4) and tested, two pre-series cavities are undergoing final welding operations (see subcomponents in Figure 5), and ten series cavities are in the beginning stages of manufacturing.

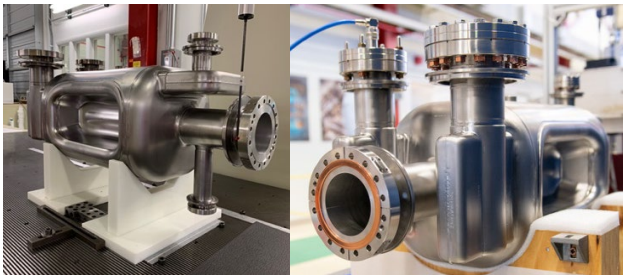

Figure 4.   First prototype built by ZRI and undergoing inspections at FNAL.

Cavities are manufactured from ultrapure RRR 300 bulk Niobium purchased and tested according to stringent technical specifications, including ultrasonic testing by immersion.

The precision required for certain components is ~0.1 mm. This required a great effort from the theoretical and practical point of view. Simulations and forming tests were necessary to design the forming, electron-beam welding, and machining equipment. Certain components such as the waveguide transitions were fabricated by bulk machining to achieve the required quality. Multiple metrological and visual tests are adopted to check the evolution of the geometries at each phase.

Issues were encountered when the deflecting poles for the pre-series cavities were formed using a new batch of material. It was later concluded that although within specifications, an intermediate batch of material was not suited for creating such an extreme formed shape. After this finding, all niobium sheets were reclassified based on their actual mechanical properties and assigned to different components based on these values.

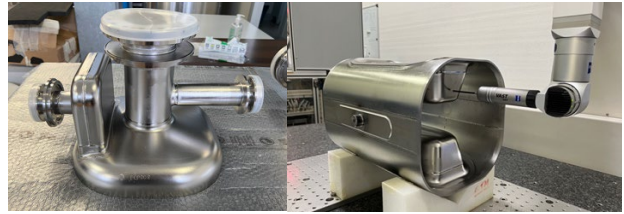

Figure 5.   Subcomponents of Pre-Series cavities during manufacturing and inspections.

Due to the stringent QA and Safety requirements set forth by CERN, great care is needed in the handling of non-conformances encountered during production at ZRI. Manufacturing data is loaded in a timely manner into the CERN EDMS and MTF system where non-conformances are managed.

## RF ANCILLARIES FABRICATION

The ancillary components for the RFD crab cavity consist of a Horizontal HOM Damper (H-HOM), Vertical HOM Damper (V-HOM) and a Field Antenna (FA); one of each is required for each dressed RFD cavity.

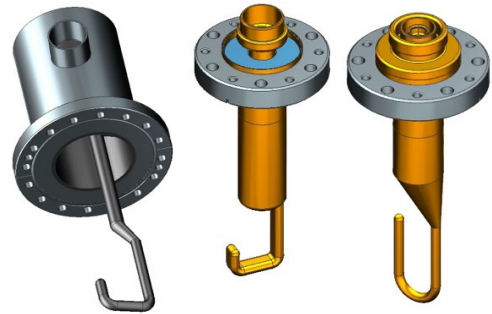

Figure 6.   H-HOM damper (left), V-HOM damper (center) and Field Pick-up antenna (right).

All RF ancillaries are under fabrication at Jlab. Three sets of prototypes have been completed and pre-series fabrication is under way.

The H-HOM damper (Figure 7Figure 6) is primarily an electron beam welded (EBW) niobium structure. A 316LN stainless steel helium jacket surrounds the inner niobium chamber. The most critical parts of the niobium structure are the Hook and Tee, which require tight mechanical tolerances to function properly. A robust system of fixturing during the EBW is required to meet RF specifications. The final QC step involves measurements on a specially designed RF test box.

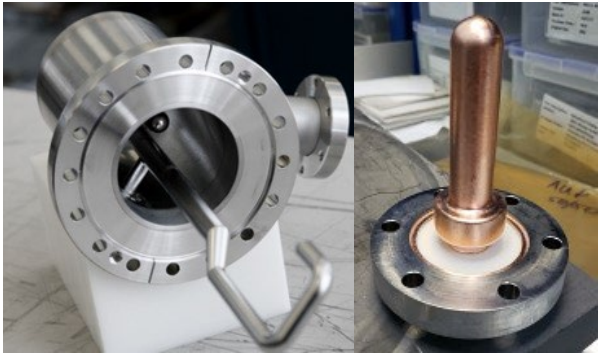

Figure 7.   H-HOM assembly (left) and feedthrough (right)

The V-HOM and FA (Figure 6 right) are designed as feedthroughs with copper probes brazed into a ceramic ring. The design proposed by Jlab makes use of a second copper ring between the ceramic and 316LN stainless steel flange. During the brazing process, the softened copper in this ring acts as a buffer to make up for the differences in thermal expansion coefficients between the ceramic and stainless steel. A common ceramic braze design is used for both components, with the two probe shapes being interchangeable.

## SURFACE PREPARATIONS

### Chemical Processing

As in most SRF cavity productions, the bulk BCP acid mixture of HF (49 %): HNO3 (69.5 %): H3PO4 (85 %) at the ratio of 1:1:2 is used. A minimum total removal of approximately 140 microns is targeted during bulk BCP.

Recently a rotating BCP tool has been developed for the RFD cavity at ANL: it is the same tool which has been used for EP on 9-cell and HWR cavities now being upgraded to process RFD cavities. The 3D model of an RFD cavity on the rotational BCP tool at Argonne National Laboratory shown in Figure 8. The rotational buffered chemical polishing of RFD cavities is performed in horizontal position, the cavity volume is filled 60% with BCP solution and the cavity rotates at 1 RPM. To keep the acid temperature between 10 and 15 degrees Celsius the temperature is kept under control using thermocouples and chilled water jets on the outer Nb surface.

Immediately after the buffered chemical polishing the cavity surface is rinsed thoroughly with DI water to remove all acid solution residue and avoid the generation of unwanted salts on the RF surface. In addition to the rotational degree of freedom the BCP tool allows the cavity to be tilted for better drainage of acid and rinsing water.

The de ionized water rinse is repeated five times to bring the measured pH of to a value in the range of 6-7.

The cavity RF surface is kept wet until the cavity receives final ultrasonic cleaning and rinsing with high-pressure water to help mitigate the problem of surface contamination adhering to the cavity surface. Ultrasonic cleaning is performed in warm water and 2% solution of Liquinox.

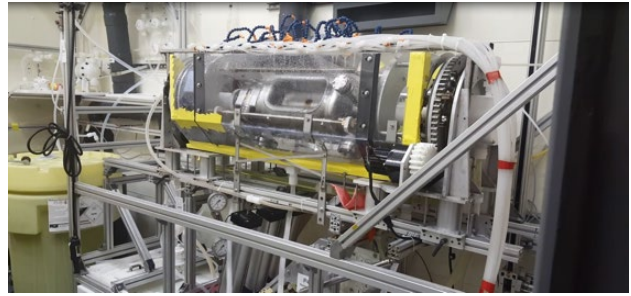

Figure 8.   RFD cavity on the rotational BCP tool at ANL, polishing in progress.

### Heat Treatment

Following the bulk BCP of the RFD cavities the cavities are heat treated at 600 C for 10 hours to degas the hydrogen absorbed during the bulk chemical processing. A residual gas analyzer is used to monitor the hydrogen evolution from the cavities during heat treatment as well as the levels of other gases. Figure 9 shows the RFD cavity in the UHV furnace, on the left, and the parameters of the heat treatment, plot on the right. The temperature has been kept at 600C for 10 hours and the RGA for mass number 2 (H) has been recorded in low $10^{-8}$ Torr at the end of the cycle, before cooling down the cavity.

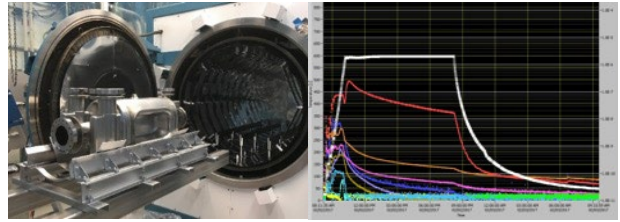

Figure 9.   RFD cavity being loaded into FNAL furnace for heat treatment (left), RGA and temperature plots during heat treatment (right).

### High Pressure Rinsing

A wand of 1.5" diameter with travel speed of 0.4"/min was used. The rotation speed of the wand was 2 rpm. During the process, the water pump pressure is 1250-1300 psi and the water temperature is ~70 F. At ANL a specific manual horizontal tool has been developed to assist an automated vertical HPR stand. The process starts with a manual external high-pressure rinse, the cavity is then rinsed manually in horizontal position, with all waveguide ports facing down. The horizontal rinse lasts 30 min per each beam port, dumping the water every 5 minutes. After that the automated vertical rinse starts, it lasts up to 3 hours. After high pressure rinsing the cavity is left to dry overnight in horizontal position, ports facing down. Once the cavity has been dried the peripherals are assembled using techniques to minimize particulate contamination which would trigger performance limiting field emission.

The RFDs are particularly complex with the multiple flange connections and orientations. Special cleanroom tooling is required to minimize contamination risk during the assembly. The cavity is then evacuated through a mass flow controller (flow rate < 30 mbar-l/sec) and leak checked to a sensitivity of 1 $e^{-10}$ mBar-l/sec$^2$ using an RGA-based leak detector.

## CRYOGENIC TESTS

### Cold Tests of Bare Cavity Prototypes

Cavities are tested first at 4.2 K and then at 2.0 K including the low power measurements to estimate the surface resistance of the cavity, achieved gradient, and field emission magnitudes. This data is recorded and captured in typical $Q_0$ and radiation vs $E_{trans}$, $E_{peak}$, $B_{peak}$ curves.

The coupling strength is determined following the intrinsic quality factor at 2.0 K and 4.2 K, with a residual resistance of 10 nΩ. An intermediate external quality factor of $5\times10^9$ is selected for the input coupler and a $2\times10^{11}$ for the pickup port.

Both prototypes manufactured by Zanon Research & Innovation, after chemical processing and preparations performed at FNAL/ANL facilities, exceeded project requirements for deflecting field and quality factor. The second prototype cavity manufactured by Zanon Research & Innovation achieved $V_t$ = 5.8 MV and $Q_0$ (4.1 MV) = 1e10 as shown in Figure 10.

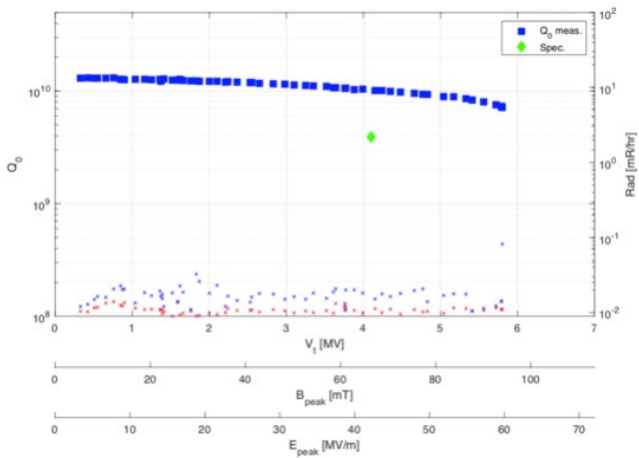

Figure 10. Cryogenic test result for bare cavity prototype #2 manufactured by Zanon Research & Innovation, processed by FNAL/ANL. Acceptance level (20% above nominal) is indicated by a green diamond.

### Cold Test with HOM Dampers Installed

A cold test for bare cavity prototype #2 with HOM dampers from Jlab was attempted at FNAL but it was inconclusive due to a leak that opened at the start of the test. The cavity was transferred to Jlab to be prepared again with HOM dampers and tested. A total of 3 cold tests have been attempted so far at Jlab most of which hindered by warm or cold leaks at various flange connections. Figure 11 shows the best result achieved with HOM dampers on this cavity. The deflecting field was slightly degraded but still exceeding requirements. The quality factor was instead below requirements.

The indication of resistive losses due to a non-superconducting material in the cavity assembly, points to an issue with the RF-shielded gasket in the flanged connection between the H-HOM and the cavity.

Measurements taken of the flange thickness in the assembled state showed that the gap between the flanges was too great to engage the 'lips' of the RF gasket, despite the connection being leak-tight. Subsequent cold tests will have the hardware in this connection torqued to a level that reduces the gap between the flanges to almost zero, ensuring that the gasket lips are crushed. Future testing will move to a new gasket with lips that are thicker.

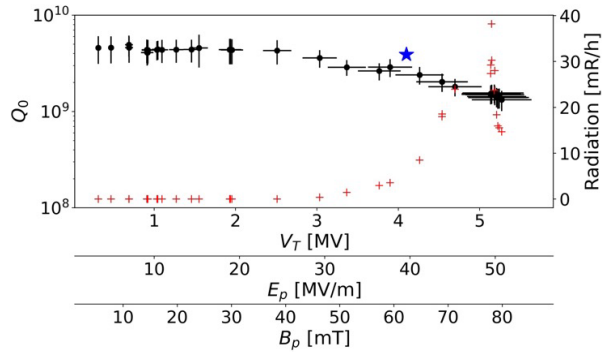

Figure 11. Test results for Prototype #2 at Jlab with HOM dampers installed. Deflecting field exceeded requirements, quality factor was below requirements.

Cold testing with an earlier iteration of the RFD cavity (LARP prototype) had similar test results, and it was addressed by using copper foil to shield the lossy stainless-steel flanges [2].

A summary of the best performing test for each cavity configuration is show in Figure 12.

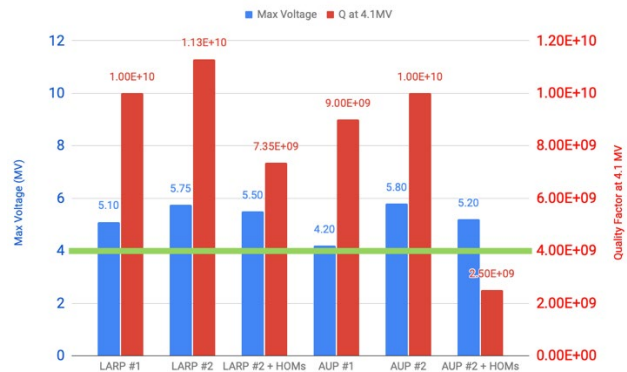

Figure 12. Summary of best results for each bare cavity built during LARP and later HL-LHC AUP. The acceptance level for Max Voltage and Quality Factor is indicated by the green line.

## SUMMARY

The AUP project is reaching peak production with cavities being fabricated in industry and HOM dampers manufactured at Jlab. Two prototype cavities were fabricated at ZRI processed at FNAL/ANL and exceeded project acceptance requirements. Chemical processing is being transferred to ZRI for the production run. Issues encountered recently at Jlab with RF losses in the H-HOM damper connection are not concerning based on past successes with the LARP R&D phase.